\documentclass[conference]{IEEEtran}
\IEEEoverridecommandlockouts

\usepackage{cite,amsmath,amssymb,graphicx,booktabs,url,xcolor,multirow}
\usepackage[hidelinks]{hyperref}
\hypersetup{
  pdftitle={Tact: A Zero-Cost, Browser-Based Pipeline for On-Demand Tactile Braille Storybooks},
  pdfauthor={Iliano Fasolino},
  pdfkeywords={braille, tactile graphics, tactile picture books, accessibility, blind children, 3D printing, language models, liblouis}
}
\graphicspath{{figures/}}

\begin{document}

\title{Tact: A Zero-Cost, Browser-Based Pipeline for On-Demand Tactile Braille Storybooks}

\author{\IEEEauthorblockN{Iliano Fasolino}
\IEEEauthorblockA{University of Milan\\
Milan, Italy\\
i.fasolino@icloud.com}}

\maketitle

\begin{abstract}
Children who learn braille have few illustrated books to read with their fingers, and almost none about the things they ask for. Existing approaches either adapt a published picture book by hand, with a specialist in the loop, or generate a tactile image without the braille page around it. We present Tact, an open system that turns a one-sentence idea into a printable page of braille with a raised picture, in Italian or English, on a consumer 3D printer, with no account and no cost beyond the plastic. Its central idea is to treat the physical page as a constraint on generation: the cell budget of a fixed tactile layout decides how long a story the language model is asked to write, and the story is measured and requested again until it fills two cards. The page itself follows requirements traced to braille standards, the tactile-graphics literature, and formative feedback from a teacher of students with visual impairments and a blind reader who printed and read a page. In an evaluation of 60 stories generated live, every story fit exactly two cards and 57 filled the second card to at least 11 of its 13 rows; a final targeted round recovered 5 of 6 retries of the remaining cases. The picture chosen matched the story's subject on 105 of 105 illustrated pages, against 63 for first-match keyword selection. Uncontracted Italian and English braille, and contracted English produced by liblouis compiled to WebAssembly, matched native liblouis cell for cell on 106 texts, and the interface showed no WCAG 2.2 AA violations. Cards are 150\,mm square on a 2\,mm plate, printed standing up, and take about 57\,g of PETG; test prints are under evaluation.
\end{abstract}


% ============================================================================
\section{Introduction}\label{sec:intro}

\subsection{Context and problem}

Reading braille is not the same as listening. Screen readers and audiobooks give a blind child access to text, but literacy, the slow and personal work of decoding words letter by letter, is built under the fingers. The decline of that skill is well documented: braille literacy among legally blind school-age children in the United States fell from about half in 1960 to under one in ten today, and nine in ten employed blind adults read braille \cite{nfb2024}. The causes are partly institutional, fewer specialist teachers per child, and partly a matter of material: there is little for a young braille reader to read.

Picture books make this gap concrete. A sighted five-year-old who asks for a story about a dragon who is afraid of the dark can have one tonight. A blind child has, at best, a small catalog of tactile picture books produced by specialists, each a fixed title written for someone else. The personal story, about the child's own dog, the thunderstorm last night, the kite that got away, does not exist in tactile form, because each page still needs a transcriber, a tactile illustrator and production equipment.

\subsection{What prior work has done}

Three lines of work address parts of this problem. First, 3D-printed tactile picture books: Stangl, Kim and Yeh used consumer 3D printing to make tactile versions of existing picture books, and studied how teachers, parents and volunteers could share the work of transcribing them \cite{stangl2014,stangl2015,kim2015}. Second, automatic or AI-assisted tactile graphics: from early systems that translated textbook figures for embossing \cite{jayant2007} to recent work that generates tactile images with generative models, for children \cite{talevision} or with blind professionals judging the result \cite{tactilenet,altcanvas,tanaka2025}. Third, braille fabricated by fused deposition modeling (FDM), where studies have measured how printed dots deviate from their design and how print orientation affects legibility \cite{zolek2024,puerta2024}.

\subsection{Limitations of prior work}

Each line solves part of the problem and leaves the rest. Tactile picture books printed in 3D adapt books that already exist, one title at a time, with people in the loop; they are not personal. Generative tactile graphics produce an image but not a page: the braille text, its layout and the composition of text and picture on one plate are left to someone else, and the image still needs expert review \cite{tanaka2025}. Studies of printed braille measure dots, not pages that combine braille with a picture. Most importantly, none of these systems connects the text to the physical space it must occupy. When text is the input, the layout has to accept whatever length arrives: a story that runs one line long spills onto a near-empty card, and one that runs short leaves blank plastic.

\subsection{Contribution}

This paper makes four contributions:
\begin{itemize}
\item \textbf{C1.} A formulation of the generated tactile picture-book page as \emph{constrained generation}, where the constraint is the page: the cell budget of a fixed tactile layout decides how many words the language model is asked for, and the system measures each story in braille cells and asks again until it fills two cards (Sections~\ref{sec:novelty} and~\ref{sec:fit}).
\item \textbf{C2.} A set of requirements for an illustrated tactile page, each traced to its source in standards, literature or formative feedback, and the layout that satisfies them (Sections~\ref{sec:method} and~\ref{sec:layout}).
\item \textbf{C3.} An open, complete system that runs in a web browser at no cost, with Italian and English uncontracted braille and contracted English (UEB Grade 2) through liblouis compiled to WebAssembly (Sections~\ref{sec:design} and~\ref{sec:impl}).
\item \textbf{C4.} A technical evaluation on stories generated live, with automatic tests that hold the braille to the reference translator at every change, and a formative print-and-read session with a blind reader (Section~\ref{sec:eval}).
\end{itemize}

Section~\ref{sec:related} reviews related work, Section~\ref{sec:novelty} states the scientific and technical novelty, Section~\ref{sec:method} the methodology and requirements, Sections~\ref{sec:design} and~\ref{sec:impl} the design and implementation, Section~\ref{sec:eval} the evaluation, and Sections~\ref{sec:discussion} and~\ref{sec:conclusion} discuss and conclude.

% ============================================================================
\section{Related Work}\label{sec:related}

\subsection{Tactile picture books and 3D printing}

Stangl, Kim and Yeh used a 3D-printed version of a classic picture book as a design probe with teachers of young blind children, finding that the objects had to be simplified and their shapes exaggerated to be recognized by touch \cite{stangl2014}. A follow-up studied how communities of transcribers, teachers and parents share the work of turning print pictures into printable tactile ones \cite{stangl2015}, and Kim and Yeh explored movable 3D-printed tactile pictures for children \cite{kim2015}. Tactipix, built on the BrailleRAP open embosser, produces illustrated children's books on paper \cite{braillerap}. All of these start from a book that already exists and rely on people to adapt it. \emph{Tact differs in} generating both the story and the page, so the book can be about whatever the child asks for.

\subsection{Recognizing tactile pictures}

Tactile pictures are hard to recognize. Theurel et al.\ asked 23 early blind children to identify pictures of eight common objects drawn with three techniques; the average identification rate was 31\%. Textured pictures were identified significantly better (36\%) than thermoformed (30\%) and raised-line pictures (27\%), and the authors attribute the weakness of raised lines to the few cues they give the fingers and to confusion between the figure and its background \cite{theurel2013}. Earlier work with blindfolded sighted adults found pictures with a uniform filled surface easier than outlines \cite{thompson2003}, and later work confirmed that textured pictures are identified well by blind children \cite{vinter2020}. Two lessons follow for a generated book: a picture should be a surface, not a line, and with recognition this hard even at best, the subjects should be simple.

\subsection{AI-generated tactile graphics}

TaleVision combines voice input, generated images and tactile output for blind children's graphic cognition, evaluated with children \cite{talevision}. TactileNet provides a dataset for training models to generate tactile graphics \cite{tactilenet}, and AltCanvas lets blind and low-vision people compose scenes with generative assistance \cite{altcanvas}. Tanaka et al.\ used image-to-image generation to design tactile pictures for inclusive picture books \cite{tanaka2023}, and in a comparison judged by totally blind professionals who supervise tactile graphics, some generated graphics were rated as highly as conventional ones, while others had lines too thin or elements overlapping, and the authors conclude that a human in the loop remains essential \cite{tanaka2025}. Chheda-Kothary et al.\ studied how blind and low-vision children want to use AI when creating visuals for their own stories \cite{chhedakothary2026}. Earlier, the Tactile Graphics Assistant automated parts of translating textbook figures for embossing \cite{jayant2007}. \emph{Tact differs in} not generating the picture at all: it chooses one from a library drawn and checked for touch, so no one has to review it, and it composes it with braille on a finished page.

\subsection{FDM braille and braille translation}

Printed dots come out smaller than designed: Żołek-Tryznowska et al.\ measured FDM dots below their design height and diameter, though still readable \cite{zolek2024}. Puerta et al.\ found that braille printed vertically or at 75 degrees was read faster and preferred over braille printed flat \cite{puerta2024}. Braille dimensions are standardized \cite{iso17049}, as are the rules of Unified English Braille \cite{ueb}, and liblouis is the reference open translator used by screen readers and braille software \cite{liblouis}. \emph{Tact differs in} combining these into a composed page, with its own measured dot geometry and translation checked against liblouis.

Table~\ref{tab:related} compares the closest systems.

\begin{table*}[t]
\caption{Comparison with the closest systems.}
\label{tab:related}
\centering
\begin{tabular}{lccccccc}
\toprule
System & Writes the text & Braille page & Tactile picture & No GPU & No expert in the loop & No cost & Usable today \\
\midrule
3D-printed picture books \cite{stangl2014,stangl2015} & No & Yes & Yes & Yes & No & Yes & Partly \\
Tactipix / BrailleRAP \cite{braillerap} & No & Yes & Yes & Yes & No & Hardware & Yes \\
TaleVision \cite{talevision} & Yes & No & Yes & Unknown & Unknown & Unknown & No \\
TactileNet \cite{tactilenet} & No & No & Dataset & No & No & Yes & No \\
AltCanvas \cite{altcanvas} & No & No & Yes & Unknown & No & Unknown & Research \\
Tanaka et al.\ \cite{tanaka2023,tanaka2025} & No & No & Yes & No & No & Unknown & No \\
Tact (this work) & Yes & Yes & Yes & Yes & Yes & Yes & Yes \\
\bottomrule
\end{tabular}
\end{table*}

% ============================================================================
\section{Scientific and Technical Novelty}\label{sec:novelty}

\subsection{Scientific novelty}

\textbf{N1: the page as a constraint on generation.} In document systems, layout adapts to text. On a tactile page it cannot: the plate is fixed, every cell is a physical feature, and a few extra words cost a whole printed card. Tact inverts the relation. The tactile layout defines a budget of braille cells; the system asks the model for the number of words that budget implies for the chosen language and braille grade; it translates the answer, counts the rows it occupies, and asks again with a corrected length until the story fills two cards. The model is treated as a stochastic component to be steered by measurement, not as a source of finished text.

\textbf{N2: traced requirements for a generated illustrated page.} The page layout follows twelve requirements, each traced to a standard, a study, or formative feedback from a teacher and a blind reader (Table~\ref{tab:req}). The resulting design, a picture in the top-right corner beside four shortened lines, a fixed left margin, no short lines inside the story, and a braille page number, is a reusable synthesis for any system that composes tactile pages automatically.

\textbf{N3: choosing a picture from the narrative.} Matching the first keyword found in a story picks whatever the story mentions first, often a detail or a figure of speech (``her heart beating''). Tact scores every named subject by where it appears in the reader's own idea, how often the page and the story mention it, and whether it is usually figurative, and picks the subject of the story rather than the first match.

\subsection{Technical novelty}

\textbf{T1.} Contracted English (UEB Grade 2) in the browser, by liblouis 3.39 compiled to WebAssembly with only the nine UEB tables embedded (about 470\,KB), loaded only when chosen. The build is usable only with a larger stack (4\,MB): with the default stack, compiling the UEB tables overflows it, and the symptom is a stream of misleading table errors.

\textbf{T2.} Uncontracted braille for Italian and English implemented in the page, following the reference tables rule for rule, and checked automatically against native liblouis at every change.

\textbf{T3.} One geometry, in millimetres, shared by the on-screen preview and the generated mesh, so the preview is the print.

\textbf{T4.} Story generation with no single point of failure and no cost: several free-tier providers raced in a staggered order behind a small serverless function that holds the keys, an in-browser model when none answers, and the reader's own words as a last resort.

% ============================================================================
\section{Methodology}\label{sec:method}

\subsection{Approach}

Tact was designed iteratively against explicit requirements. Requirements came from three sources: braille and accessibility standards, the literature in Section~\ref{sec:related}, and formative feedback from two people with complementary expertise. A teacher of students with visual impairments reviewed the website and the page design. A blind reader, an experienced braille user who works with tactile graphics, printed a page standing upright on a consumer printer and read it, and reported on the dots, the picture, the blank space, the plate and the braille grade. This is formative design feedback, not a user study, and the evaluation in Section~\ref{sec:eval} is reported with that scope.

\subsection{Requirements}

Table~\ref{tab:req} lists the requirements, their sources and the design element that satisfies each. Some deserve a word. The teacher advised placing pictures at the beginning or end of a story so that braille lines are not cut short in the middle (R2). The blind reader found that a small picture centred in a band left an ``ocean'' of blank plastic, that no braille appeared until a third of an upright print was done, and that an outline picture could not be recognized (R4 to R6); the reader also asked for contracted braille, since many children learn contractions from the start (R8), and found that a thin page printed upright buckled (R12).

\begin{table*}[t]
\caption{Requirements, their sources, and the design element that satisfies each.}
\label{tab:req}
\centering
\begin{tabular}{llll}
\toprule
\# & Requirement & Source & Design element (section) \\
\midrule
R1 & Dot and cell dimensions within ISO 17049, oversized for FDM & Standard \cite{iso17049}, measurement \cite{zolek2024} & Dot geometry (\ref{sec:fab}) \\
R2 & No short braille lines in the middle of a story & Teacher & Picture beside the first four lines only (\ref{sec:layout}) \\
R3 & Every line starts at the same left margin & Braille reading practice & Picture on the right, not the left (\ref{sec:layout}) \\
R4 & No large blank areas; each card is full & Blind reader & Story fitted to two cards (\ref{sec:fit}) \\
R5 & Braille reached early in an upright print & Blind reader & Picture at the top, lines from the top (\ref{sec:layout}) \\
R6 & Recognizable picture: filled, flat view, simple subject & Blind reader, \cite{theurel2013,thompson2003} & Silhouette library (\ref{sec:pictures}) \\
R7 & Cards can be ordered without sighted help & Braille book practice & Braille page number (\ref{sec:layout}) \\
R8 & Contracted English available; uncontracted by default & Blind reader, \cite{ueb} & UEB Grade 2 option (\ref{sec:braille}) \\
R9 & No GPU, no payment, no account & Equity of access & Zero-cost architecture (\ref{sec:impl}) \\
R10 & Italian and English as equal languages & Context of use & Bilingual pipeline \\
R11 & Accessible interface, voice input & WCAG 2.2 \cite{wcag22} & Interface (\ref{sec:interaction}) \\
R12 & A plate that stays flat and steady when printed upright & Blind reader & 2\,mm plate (\ref{sec:fab}) \\
\bottomrule
\end{tabular}
\end{table*}

% ============================================================================
\section{System Design}\label{sec:design}

\subsection{Pipeline}

\begin{figure}[t]
\centering
\includegraphics[width=\columnwidth]{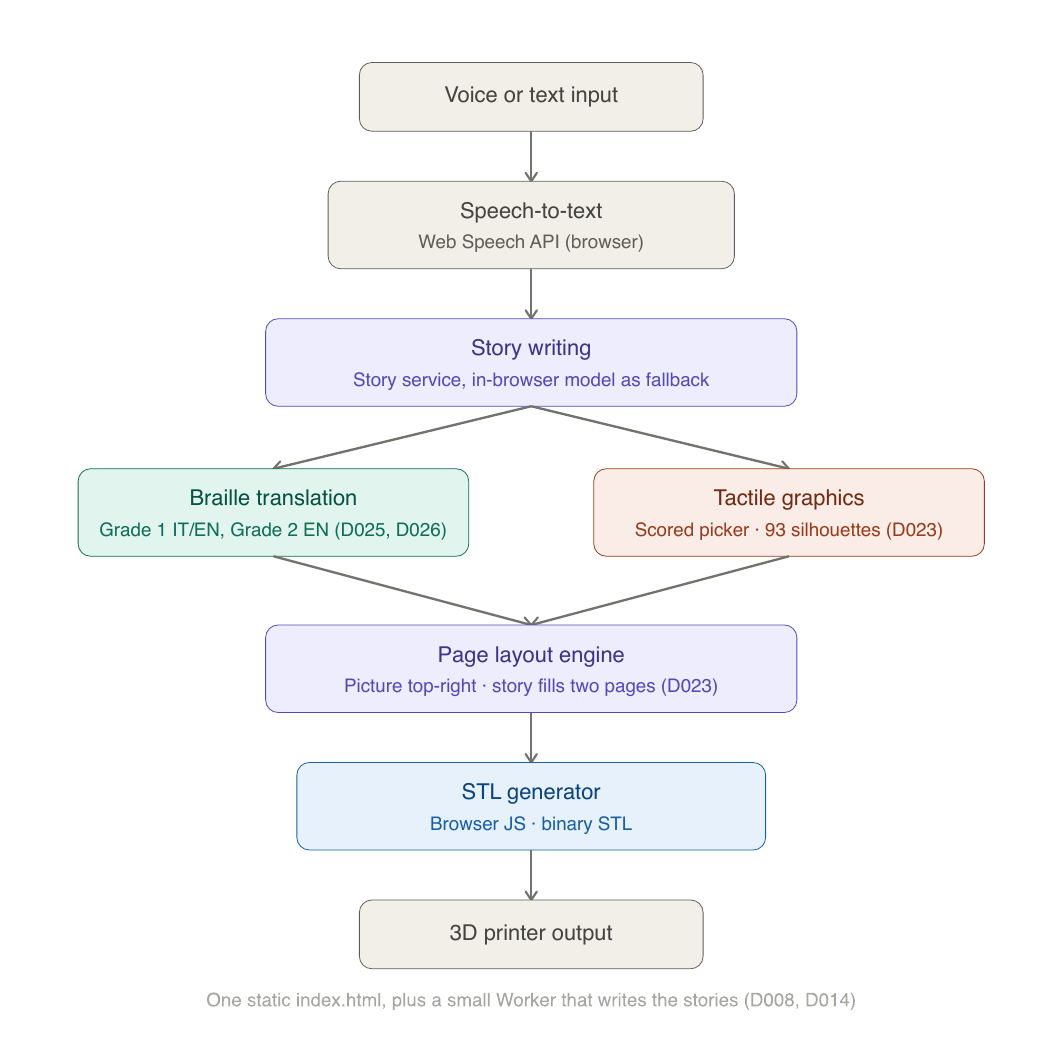}
\caption{The pipeline. The story is written by a small serverless function that races free-tier language-model providers, with an in-browser model as fallback; braille, layout, picture choice and mesh generation run in the page.}
\label{fig:pipeline}
\end{figure}

A reader or helper types or speaks one sentence. Tact writes a story about it, translates the story into braille, splits it into cards, chooses a picture, and shows a true-to-scale preview with a downloadable mesh for each card (Fig.~\ref{fig:pipeline}). Everything after the story is computed in the browser.

\subsection{Page layout}\label{sec:layout}

\begin{figure}[t]
\centering
\includegraphics[width=\columnwidth]{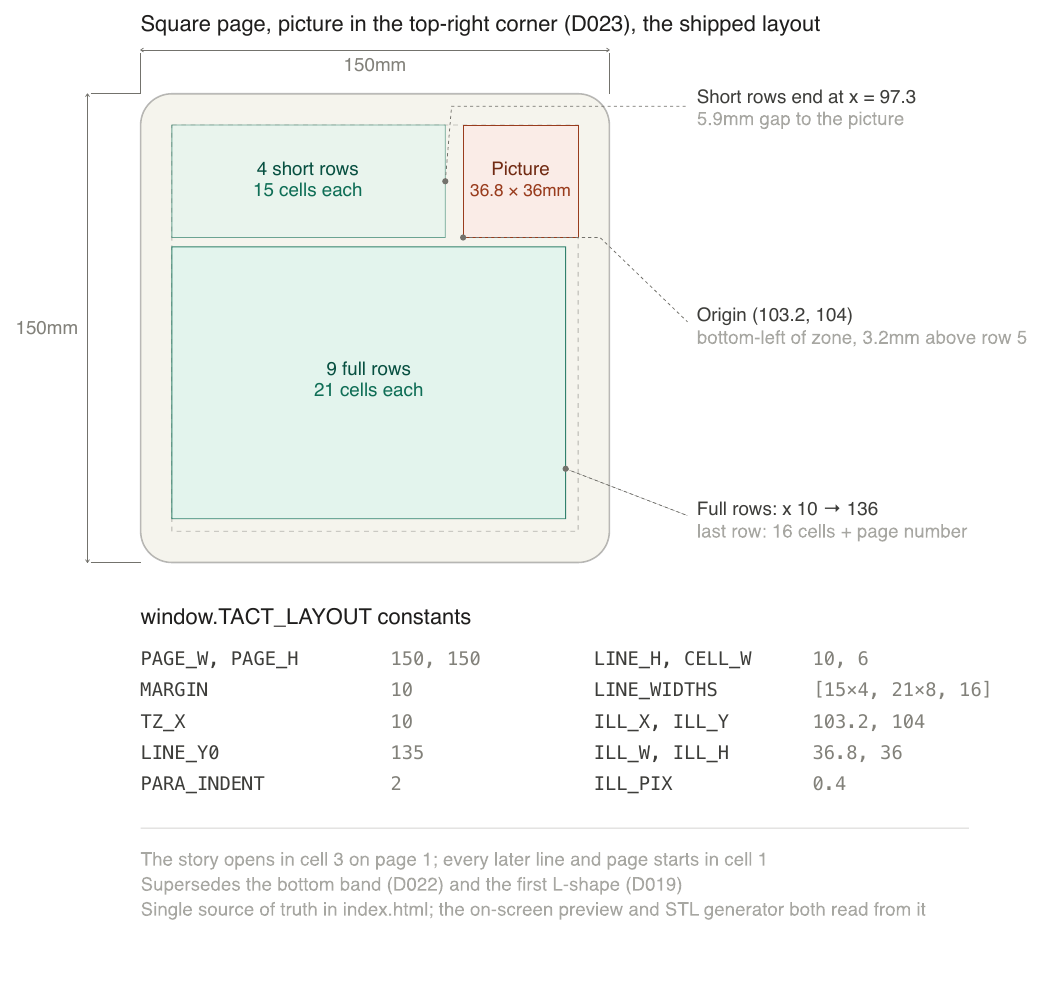}
\caption{Card geometry in millimetres: 13 rows, the first four shortened to 15 cells beside the 36.8 by 36\,mm picture zone, eight full rows of 21 cells, and a last row of 16 cells followed by the braille page number.}
\label{fig:geometry}
\end{figure}

A card is a square plate of 150 by 150\,mm, which fits the bed of common consumer printers. It holds 13 braille rows on a 10\,mm line pitch with a 6\,mm cell pitch and 2.5\,mm between dots, within ISO 17049 \cite{iso17049}. The picture occupies a 36.8 by 36\,mm zone in the top-right corner; the first four rows beside it hold 15 cells, the next eight hold 21, and the last holds 16 cells of text followed by the page number, right-aligned at cell 21 (Fig.~\ref{fig:geometry}). An illustrated card carries 244 cells of text, a text-only card 268. The story opens in cell 3, the conventional paragraph indent.

\begin{figure}[t]
\centering
\includegraphics[width=\columnwidth]{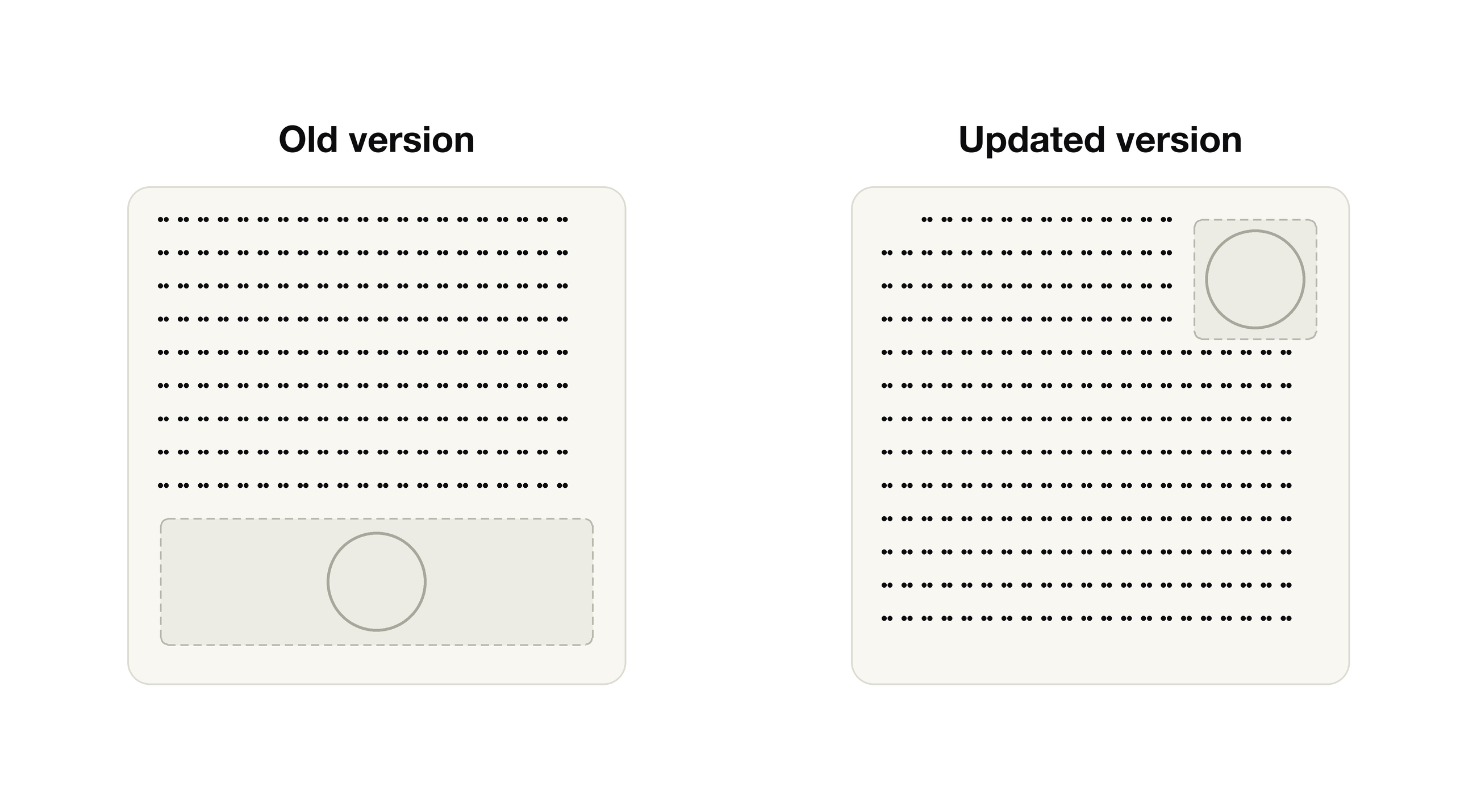}
\caption{Two placements of the picture on the same card. Left: a band across the bottom keeps every line full but costs four rows and leaves blank space beside a small picture. Right: the top-right corner shortens only the first four lines and keeps the left margin fixed.}
\label{fig:band}
\end{figure}

The corner placement answers several requirements at once. Compared with a picture in a band across the bottom (Fig.~\ref{fig:band}), it shortens four lines to 71\% of their width instead of removing rows, reaches the braille from the first line of an upright print (R5), and leaves no blank sides (R4). The blind reader suggested the top-left corner, with the opening lines indented beside the picture. The top-right corner was chosen instead because it keeps every line starting at the same left margin, where a reader's hand returns at the end of each line (R3), and follows the picture-book convention of text beginning at the left. The page number sits where braille books put it, so a reader can order the cards alone (R7).

\subsection{Fitting the story to the page}\label{sec:fit}

The page cannot grow, so the story must fit it. A story that fills two cards needs about 25 of the 26 available rows; measured on stories from the live service, a word takes about 0.345 rows in English and 0.37 in Italian (capital signs, accents and word wrapping included), and contracted English packs about 0.78 of the uncontracted cells. The first request therefore asks for 70 English words, 66 Italian words, or 88 contracted English words, and a second request in parallel asks for 10\% more.

The model does not count words precisely, so each answer is translated and measured. Fig.~\ref{fig:loop} summarizes the procedure. A story is accepted when it fills exactly two cards and the second is at least 84\% full, 11 of its 13 rows. Otherwise a correction is computed from the answer's own density (rows per word) and from how far the model missed the last request, and two new lengths are requested, the computed one and 8\% less, since a one-row overshoot costs a third card. After three such rounds, a last round aims from the typical drift across all answers so far rather than the last one, and spreads three requests around that aim. The best story by a simple score (two cards first, then the fuller second card) is always kept, so each round can only improve the page. Requested lengths are clamped to 45 to 110 words, and a story costs at most eleven requests, usually two.

Contracted braille is loaded before the first length is chosen, because a story measured in uncontracted cells looks longer than it will print.

\begin{figure}[t]
\centering
\fbox{\begin{minipage}{0.94\columnwidth}\small
\textbf{Input:} idea $q$, language $\ell$, grade $g$.\\
1. Ask for $w_0(\ell,g)$ words and, in parallel, $1.1\,w_0$; keep the better story $s$.\\
2. \textbf{repeat} up to 3 times \textbf{while} not $\mathit{good}(s)$:\\
\hspace*{1em} $r \leftarrow$ rows of $s$ in braille; $d \leftarrow \mathit{words}(s)/r$;\\
\hspace*{1em} $b \leftarrow \mathit{clamp}(\mathit{words}(s)/w_{\mathit{asked}}, 0.6, 1.4)$;\\
\hspace*{1em} $w \leftarrow 25\,d / b$; ask for $w$ and $0.92\,w$; keep the best.\\
3. \textbf{if} not $\mathit{good}(s)$: $b \leftarrow$ median drift of all answers;\\
\hspace*{1em} $w \leftarrow 25\,d/b$; ask for $w$, $1.08\,w$, $0.93\,w$; keep the best.\\
$\mathit{good}(s)$: two cards, second card at least 11 of 13 rows.
\end{minipage}}
\caption{Fitting the story to two cards.}
\label{fig:loop}
\end{figure}

The editorial brief, which stays on the server, asks for a story with a clear arc in short sentences, and favours touch, sound, warmth and movement over colour and sight.

\subsection{Braille translation}\label{sec:braille}

Uncontracted braille needs no dictionary, but it is not a plain letter-for-cell table: Italian and English mark capitals, quotation marks and brackets differently, and both depend on context, such as a capitalized word, an opening or closing quote, or a letter directly after a number. Tact implements these rules for both languages in the page, following the liblouis tables \texttt{en-ueb-g1} and \texttt{it-it-comp6} \cite{liblouis}: for example, the Italian capital sign is dots 4-6 and the English one dot 6, the English question mark is dots 2-3-6 and the Italian one dots 2-6, and a run of three or more fully capitalized English words takes a capitals passage indicator and a terminator. Typographic quotes, apostrophes and ellipses produced by language models are folded onto the forms the tables know, and a long dash stays a dash.

Contracted English uses liblouis itself (T1), because contraction rules depend on syllables and word position and a hand-written table would teach a child wrong braille. It is offered under the text box for English, uncontracted remains the default, and if the translator fails to load the story is printed uncontracted and the page says so (R8).

\subsection{Picture library and choice}\label{sec:pictures}

\begin{figure}[t]
\centering
\includegraphics[width=\columnwidth]{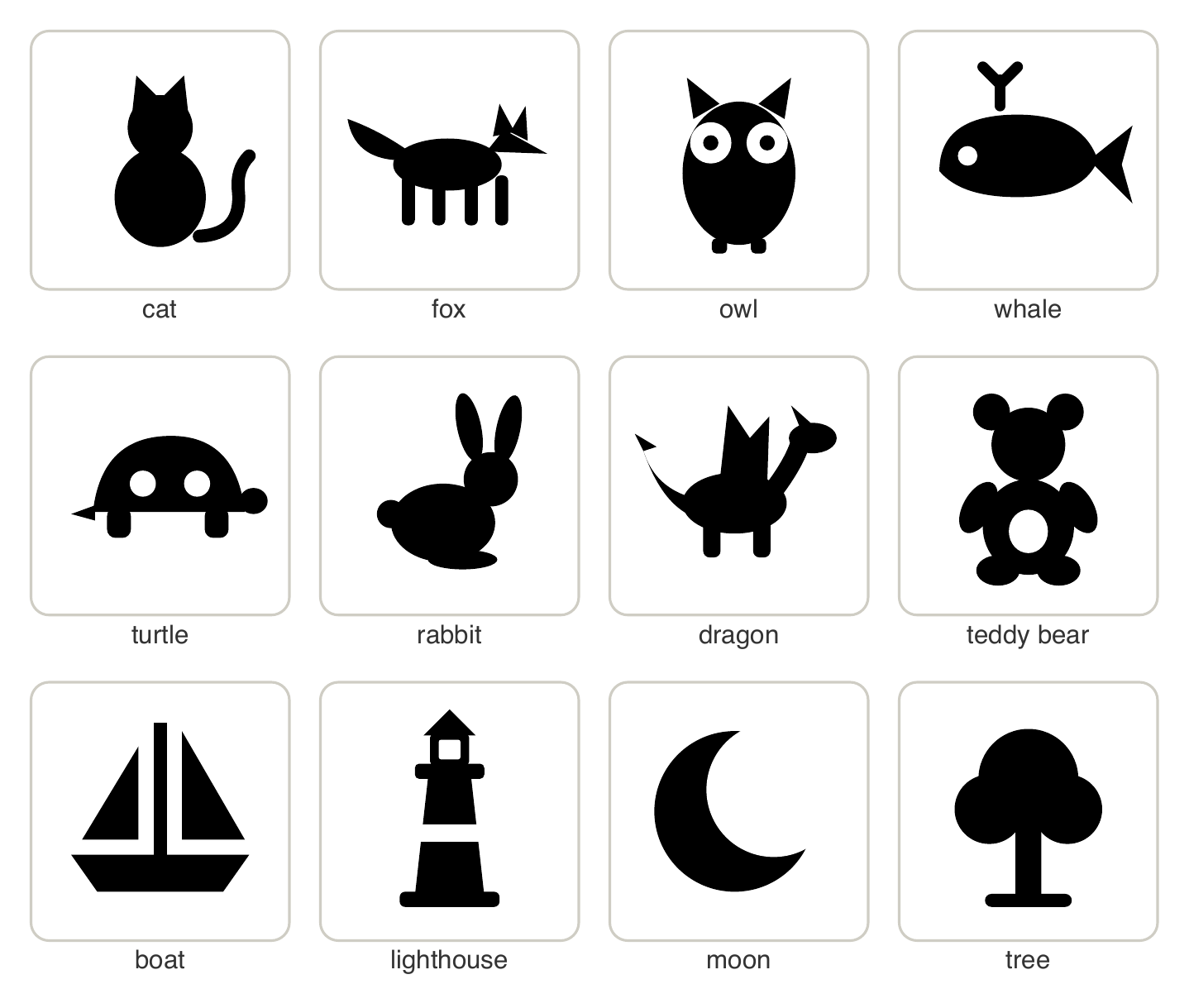}
\caption{Twelve of the 93 silhouettes in the picture library: filled surfaces in a flat side or front view, with thick parts and wide gaps at print size.}
\label{fig:shapes}
\end{figure}

The picture library holds 93 hand-drawn silhouettes of common story subjects: animals, fairy-tale figures, everyday objects, weather and sky. Each is a filled surface in a flat side or front view, with no perspective, strokes of at least 2.5\,mm and gaps of at least 3\,mm at print size (R6). The relief is built at 0.4\,mm per pixel, the width of a standard nozzle, cropped to the drawing and fitted into the corner zone, and stands 1.0\,mm above the plate, 0.15\,mm above the dots, so the fingers can tell picture from text. The picture size suits simple subjects; the library contains only subjects that read at that size, which the recognition rates in \cite{theurel2013} argue for. Fig.~\ref{fig:shapes} shows twelve of them.

Words are matched as whole words in each language, with plurals and diminutives (726 words in all), and no word belongs to two shapes. Each named shape receives a score:
\begin{equation}
S = \max(10,\,20 - 2p) \cdot [p \text{ exists}] + 2\,n_{\mathit{page}} + n_{\mathit{story}},
\end{equation}
where $p$ is the position of the first matching word in the reader's idea, and $n_{\mathit{page}}$ and $n_{\mathit{story}}$ count mentions on the page and in the whole story. Shapes whose words are mostly figurative (heart, star, pirate hat) are weighted by 0.3, and ties go to the earlier mention on the page. When the idea names a particular kind of a shape, the story's general word counts for it, so ``a teddy bear'' still shows the teddy bear when the story only says ``the bear''. Each illustrated card is scored on its own text, falling back to the story's main subject.

\subsection{Interaction}\label{sec:interaction}

The interface accepts speech through the browser's speech recognition, works with screen readers, keyboard and touch, and uses short synthesized sounds to mark listening, waiting and results. It is bilingual, respects the system setting for reduced motion, and offers each card as a mesh and the whole story as a braille ASCII file (BRF) that a teacher can open in braille software or send to an embosser.

% ============================================================================
\section{Implementation and Fabrication}\label{sec:impl}

\subsection{A zero-cost architecture}

The application is one static HTML page, served by a free static host. The only server component is a small serverless function on a free tier that holds the providers' keys, keeps the editorial brief out of reach of callers, accepts requests only from the site's own origins, and counts requests exactly per address (20 a minute) and globally (240 a minute). It races five models from three free-tier providers, two of them with two models each, whose allowances are separate: the preferred model starts at once, the next starts if it has not answered after a short grace period, the first story wins, and a model that fails is skipped for a cooldown. Models that run long or short of a requested length are asked for proportionally less or more. If none answers, the page runs a small language model in the browser through WebGPU \cite{webllm}, choosing the largest that the device can hold up to 3B parameters; if that is impossible too, the reader's own words become the page. Table~\ref{tab:cost} lists every step and its cost.

\begin{table}[t]
\caption{Every step of the pipeline and its cost.}
\label{tab:cost}
\centering
\begin{tabular}{p{2.2cm}p{3.9cm}p{0.8cm}}
\toprule
Step & Technology & Cost \\
\midrule
Speech input & Browser speech recognition & Free \\
Story & Serverless function, free-tier providers; in-browser model as fallback & Free \\
Braille & In-page rules; liblouis in WebAssembly for Grade 2 & Free \\
Pictures & Local library of 93 silhouettes & Free \\
Layout, mesh & Computed in the page & Free \\
Hosting & Static host, serverless free tier & Free \\
Printing & Consumer FDM printer & Plastic, about 57\,g per card \\
\bottomrule
\end{tabular}
\end{table}

\subsection{Privacy}

With the hosted path, one sentence leaves the device, and the function keeps no record of it; the per-address counter holds only request times for one minute. With the in-browser path nothing leaves the device. There are no accounts.

\subsection{Mesh generation}

Each card becomes a binary STL file generated in the page. The plate is a square with 10\,mm rounded corners, 2\,mm thick, with a 0.4\,mm chamfer on its bottom edge, which absorbs the spread of the first printed layer. Each dot is a dome on a short cylinder, 1.6\,mm in diameter and 0.85\,mm tall as designed; the picture is a set of flat-topped blocks, one per run of pixels, merged across rows. The same geometry, defined once in millimetres, draws the on-screen preview (T3).

\subsection{Fabrication}\label{sec:fab}

\begin{figure}[t]
\centering
\includegraphics[width=\columnwidth]{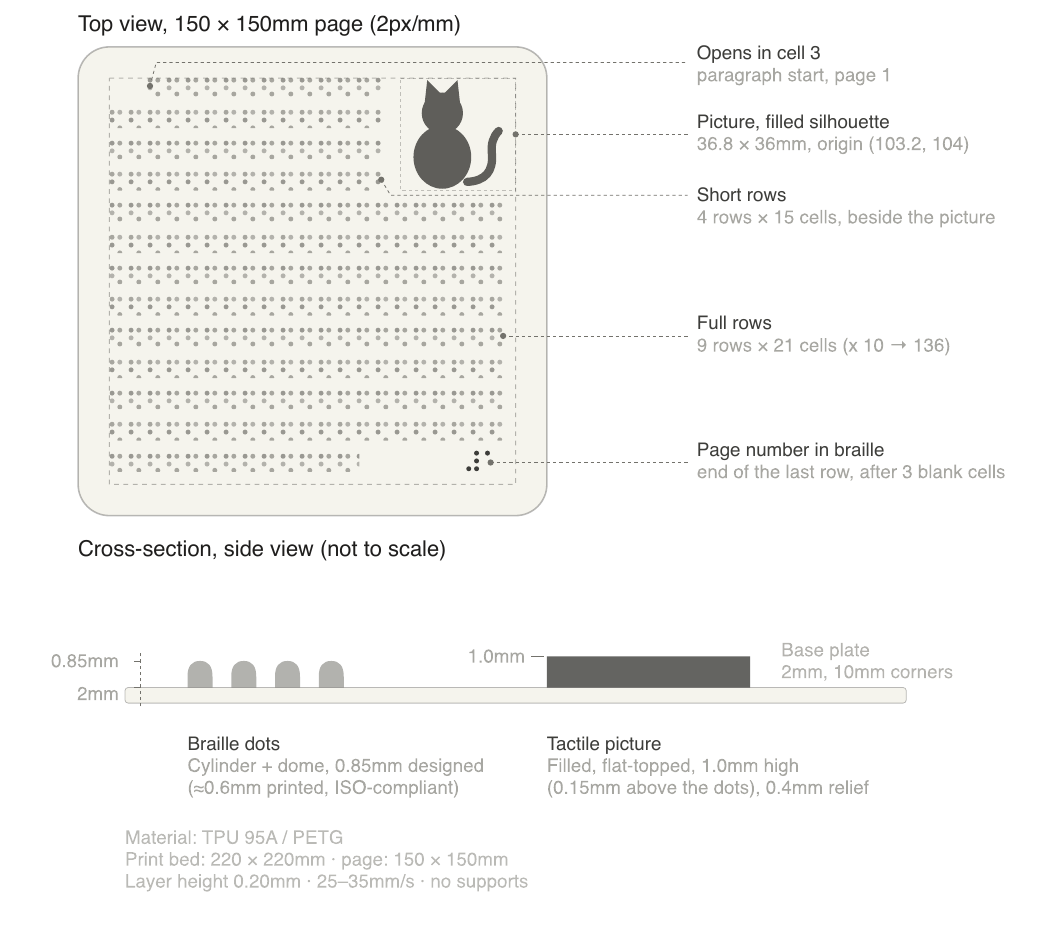}
\caption{A card: top view with the silhouette in the top-right corner, and a cross-section comparing domed braille dots with the flat-topped picture relief, on a 2\,mm plate.}
\label{fig:mockup}
\end{figure}

FDM printers produce dots smaller than designed. On a reference print, dots designed at 0.85\,mm tall and 1.6\,mm wide measured about 0.60\,mm and 1.3\,mm, within the ISO 17049 range; the design is oversized accordingly (R1). Fig.~\ref{fig:mockup} shows the card and its cross-section. Printed standing on its bottom edge, a card gives the roundest dots, consistent with \cite{puerta2024}, but needs a plate stiff enough to stay flat and steady during the print; stiffness grows with the cube of thickness, so a 2\,mm plate is 64 times stiffer than a 0.5\,mm one. A card therefore has a 2\,mm plate and is printed standing up; it takes about 57\,g of PETG, and the dots rise 0.85\,mm and the picture 1.0\,mm above the plate. Test prints are in development and under evaluation.

\subsection{An independent implementation}

Tact is open source under the MIT license. Francesco Giuliani independently rebuilt it as a production web application, tatto.dev, a fork that precompiles the interface, ports the braille tables and mesh generation to typed modules with unit tests on the braille mappings, reaches a hosted language model through a server route, and reduces the first-load JavaScript by about 94\% \cite{tattodev}. That work showed the value of a hosted generation path in front of the in-browser model, and it is evidence that the design can be reproduced and extended by others.

\subsection{Engineering challenges}

Three problems shaped the implementation; each is reported with its cause and the current solution.

\emph{Contracted braille in the browser.} A WebAssembly build of liblouis failed with errors that pointed at the UEB tables. An earlier report of this work attributed the failure to an unsupported opcode \cite{tactv2}; the actual cause was a stack overflow while compiling the tables, and the build works with a 4\,MB stack, with output identical to native liblouis (Section~\ref{sec:evalbraille}).

\emph{Free tiers move.} Over the life of the service, providers retired model names overnight, and one key could list models it was not allowed to use. A provider is therefore admitted to the race only after a live test, the race tolerates any member failing, and diagnostic endpoints test each provider in isolation. A check on 28 and 29 October 2026 found all three backups unavailable while the preferred model kept answering: one provider had retired the configured model, another had spent the free allowance of its model, and a reasoning model spent its 600-token output budget thinking and returned empty stories. The fixes were specific: the provider's newer model, a second and smaller model on the same key as the spent one, with its own allowance, and low reasoning effort with a larger budget. Models also differ in how closely they follow a requested length: the smaller model wrote 1.4 to 1.6 times the request, and asked for 70\% of it, it landed at 1.1 to 1.4 times in four probes. After these changes four of the five models wrote stories in both languages; the fifth is still limited by its provider's allowance, and since it is refused in about 0.1\,s it costs the race nothing (Section~\ref{sec:evalservice}).

\emph{Uncontracted braille is not a table.} A single character table shared by both languages printed Italian capitals with the English sign and the English question mark with the Italian cell. The current rules follow the reference tables per language, and an automatic test at every change compares them with liblouis cell for cell.

% ============================================================================
\section{Evaluation}\label{sec:eval}

All measurements were taken on the deployed system in October 2026, with requests spaced in time. The data and scripts are released with the paper.

\subsection{Braille correctness}\label{sec:evalbraille}

The test corpus contains the 60 generated stories of Section~\ref{sec:evalfit} (40 English, 20 Italian) and 46 hand-written sentences exercising capitals, capitalized words and passages, numbers with separators, quotation marks, brackets, accents, dashes and ellipses. Expected output was produced by native liblouis 3.39.0 with \texttt{en-ueb-g1}, \texttt{it-it-comp6} and \texttt{en-ueb-g2}. Tact's translator, loaded from the shipped page, matched on every text (Table~\ref{tab:braille}); the test runs automatically at every change to the code.

\begin{table}[t]
\caption{Tact's braille against native liblouis 3.39.0.}
\label{tab:braille}
\centering
\begin{tabular}{lrrr}
\toprule
Translation & Texts & Identical & Cells \\
\midrule
Uncontracted English & 67 & 67 & 19,518 \\
Uncontracted Italian & 39 & 39 & 8,262 \\
Contracted English (WebAssembly) & 67 & 67 & 15,289 \\
\bottomrule
\end{tabular}
\end{table}

\subsection{Fitting the page}\label{sec:evalfit}

Twenty ideas, each with a subject from the library and often a second object (a castle, the moon, a key), were run in three conditions, Italian, English uncontracted and English contracted, for 60 stories generated live, one every three minutes through the real interface. Every story filled exactly two cards; none spilled onto a third. Table~\ref{tab:fit} reports the second card.

\begin{table}[t]
\caption{Second-card fill, requests and time, 60 live stories (three rounds of correction).}
\label{tab:fit}
\centering
\begin{tabular}{lccc}
\toprule
 & Italian & English & English G2 \\
\midrule
Second card $\geq$ 11/13 rows & 18/20 & 20/20 & 19/20 \\
Rows: 13 / 12 / 11 / $\leq$10 & 7/9/2/2 & 7/10/3/0 & 5/7/7/1 \\
Median requests & 2 & 2 & 2 \\
Median time & 1.4\,s & 1.4\,s & 2.1\,s \\
Words (min / median / max) & 54/68/74 & 68/73/85 & 77/96/104 \\
\bottomrule
\end{tabular}
\end{table}

Forty-two stories needed only the first two requests. End-to-end time from submission to result had a median of 1.8\,s and a 90th percentile of 4.3\,s. Contracted stories carry about 30\% more words and still fill the page, which confirms measuring in the target grade. The three stories left short had used all eight requests of the three rounds. The last round of Section~\ref{sec:fit} was then evaluated on exactly these cases: repeated twice for each, it filled the second card in 5 of 6 tries, and in the sixth the earlier story was kept. A fresh run of 9 stories through the interface, three per condition, filled every second card (11 to 13 rows) with two or four requests, and never needed the last round.

How long the model writes compared with what it is asked shows why measurement is needed. Across 175 answers the median ratio of words received to words asked was 0.96 in Italian, 1.01 in English and 0.93 in contracted English, but single answers ranged from 0.66 to 1.34, and only 59\%, 84\% and 68\% landed within 15\% of the request. The model is right on average and unreliable per story.

\subsection{Picture choice}

On the same 60 stories, the chosen picture matched the expected subject on the first card of 60 stories and on 105 of 105 illustrated cards. Applied to the same texts, first-match selection in table order scored 46 of 60 first cards and 63 of 105 cards: the turtle became the rabbit it races, the lighthouse the ship it guides, the squirrel the key it hides, and many second cards showed a heart from ``her heart beat fast''. Choosing from the reader's idea alone scored 57 of 60 and failed when the idea used a word the story did not (a pirate cat became a pirate hat). A separate set of 50 ideas written with diminutives, plurals and a second object (``un volpacchiotto che cerca la luna'', ``three kittens who find a key'') all chose the subject.

\subsection{Meshes and generation service}\label{sec:evalservice}

Generated meshes were valid on every card checked: byte length equal to 84 plus 50 bytes per triangle, no degenerate triangles, a 150 by 150\,mm footprint, and 37,000 to 42,000 triangles per card. Of 180 story requests, 179 succeeded; one received a rate-limit response while its parallel request answered. The median request latency was 0.85\,s. All successful requests were won by the preferred model, so this run does not measure the backups. They were probed one by one after the fixes of Section~\ref{sec:impl}, asking for the page's lengths in Italian and English: four of the five models wrote a story in both languages, the preferred one in 0.3\,s and the backups in 1.2 to 2.7\,s, with lengths from 0.89 to 1.39 times the request; the fifth was refused by its provider's allowance.

\subsection{Accessibility}

An automated audit with axe-core 4.10.2 against WCAG 2.2 AA found no violations on the home screen in both languages, a result screen and the information page in both languages. Color contrast, which the tool could not compute because of decorative layers, was computed by hand: 18.9:1 for body text and at least 5.2:1 for the smallest labels.

\subsection{Printing and reading}

Test prints of the 2\,mm card, printed standing up, are in development and under evaluation.

In the formative session, the blind reader printed a card upright and read it. The dots of the upright print were good and the braille read well; the reader's remaining observations, blank space, a picture that could not be recognized, braille reached late in the print, a thin plate that buckled, and the wish for contractions, became requirements R4, R6, R5, R12 and R8, and are addressed by the design in Section~\ref{sec:design}.

\subsection{Scope of the evaluation}

The evaluation establishes that the system produces correct braille, well-filled cards and fitting pictures, reliably and quickly; it does not measure reading or learning outcomes in children, which require a study with children and their teachers.

% ============================================================================
\section{Discussion}\label{sec:discussion}

\textbf{When the page drives the text.} Treating the page as a constraint turns the language model into a component to be calibrated rather than an author to be trusted. Its errors are not in content but in length, and they are large for single answers even when the average is right. Measuring in the target braille grade and asking again is cheap, two requests for most stories and a second of waiting, and it removes the near-empty card that a fixed-length request produces one time in several.

\textbf{Choosing rather than generating pictures.} A library limits variety: a story about a submarine gets the boat or the fish, or the star when nothing matches. In return every picture has been drawn for touch and needs no review, which recent work on generated tactile graphics still requires \cite{tanaka2025}. Scoring by the reader's idea and the narrative, rather than by the first word found, is what makes a fixed library feel chosen.

\textbf{Both braille grades.} Offering contracted English alongside uncontracted braille leaves the choice to the teacher, who knows how the child is learning. Holding both to the reference translator with an automatic test keeps the choice honest as the code changes.

\textbf{Cost and access.} The system needs a browser and access to a consumer FDM printer, increasingly common in schools and libraries. Generation costs nothing to the reader and little to run; the material cost is the plastic, about 57\,g per card. Free tiers are not guaranteed, and the design accepts that: any provider can fail, the browser can write the story, and the reader's words always make a page.

\textbf{Ethics.} Feedback was given voluntarily by people who agreed to be acknowledged by name. The hosted path sends one sentence and keeps no record of it, and no account exists to hold a child's stories.

% ============================================================================
\section{Conclusion}\label{sec:conclusion}

Tact makes a personal tactile picture book from one sentence: a story written to fill two braille cards, a picture chosen for touch from what the reader asked for, and a mesh ready for a consumer printer, at no cost. Treating the physical page as a constraint on generation filled the second card in 57 of 60 live stories with three rounds of correction, and a final round recovered most of the rest; scoring pictures by the narrative chose the right subject on every illustrated card; and the braille, uncontracted Italian and English and contracted English, matches the reference translator cell for cell. The requirements behind the page, traced to standards, studies and the people who read it, are a reusable basis for any system that composes tactile pages automatically.

\section*{Acknowledgment}

The author thanks Chris Bischke, Ph.D., TVI, DT/V (University of Utah and Utah State University), for reviewing the website and for the advice to place pictures at the beginning or end of a story. The author thanks Chancey Fleet, Blind technologist with a graphics and braille lab at the New York Public Library and board member of the American Action Fund for Blind Children and Adults, for printing and reading a card and for feedback on the layout, the picture, the plate and contracted braille. The author thanks Francesco Giuliani, whose independent implementation tatto.dev showed the value of a hosted generation path.

% ============================================================================


\begin{thebibliography}{99}

\bibitem{nfb2024} National Federation of the Blind, ``The braille literacy crisis in America.'' [Online]. Available: https://nfb.org/braille-illiteracy-crisis

\bibitem{stangl2014} A. Stangl, J. Kim, and T. Yeh, ``3D printed tactile picture books for children with visual impairments: A design probe,'' in \emph{Proc. Interaction Design and Children (IDC)}, 2014, pp. 321-324, doi: 10.1145/2593968.2610482.

\bibitem{stangl2015} A. Stangl, C.-L. Hsu, and T. Yeh, ``Transcribing across the senses: Community efforts to create 3D printable accessible tactile pictures for young children with visual impairments,'' in \emph{Proc. ACM SIGACCESS Conf. Computers and Accessibility (ASSETS)}, 2015, pp. 127-137, doi: 10.1145/2700648.2809854.

\bibitem{kim2015} J. Kim and T. Yeh, ``Toward 3D-printed movable tactile pictures for children with visual impairments,'' in \emph{Proc. ACM CHI Conf. Human Factors in Computing Systems}, 2015, pp. 2815-2824, doi: 10.1145/2702123.2702144.

\bibitem{braillerap} BrailleRAP Project, ``BrailleRAP: An open-source DIY braille embosser.'' [Online]. Available: https://www.braillerap.org/en/

\bibitem{theurel2013} A. Theurel, A. Witt, P. Claudet, Y. Hatwell, and E. Gentaz, ``Tactile picture recognition by early blind children: The effect of illustration technique,'' \emph{J. Experimental Psychology: Applied}, vol. 19, no. 3, pp. 233-240, 2013, doi: 10.1037/a0034255.

\bibitem{thompson2003} L. J. Thompson, E. P. Chronicle, and A. F. Collins, ``The role of pictorial convention in haptic picture perception,'' \emph{Perception}, vol. 32, no. 7, 2003, doi: 10.1068/p5020.

\bibitem{vinter2020} A. Vinter, O. Orlandi, and P. Morgan, ``Identification of textured tactile pictures in visually impaired and blindfolded sighted children,'' \emph{Frontiers in Psychology}, vol. 11, 2020, doi: 10.3389/fpsyg.2020.00345.

\bibitem{talevision} H. Wu, H. Yang, F. Chang, D. Zhu, and Z. Liu, ``AI-generated tactile graphics for visually impaired children: A usability study of a multimodal educational product,'' \emph{Int. J. Human-Computer Studies}, 2025, doi: 10.1016/j.ijhcs.2025.103525.

\bibitem{tactilenet} A. Khan, A. Choubineh, M. A. Shaaban, A. Akkasi, and M. Komeili, ``TactileNet: Bridging the accessibility gap with AI-generated tactile graphics for individuals with vision impairment,'' arXiv:2504.04722, 2025.

\bibitem{altcanvas} S. Lee, M. Kohga, S. Landau, S. O'Modhrain, and H. Subramonyam, ``AltCanvas: A tile-based editor for visual content creation with generative AI for blind or visually impaired people,'' in \emph{Proc. ASSETS}, 2024.

\bibitem{tanaka2023} K. Tanaka, T. Fushimi, A. Tsutsui, and Y. Ochiai, ``Text to haptics: Method and case studies of designing tactile graphics for inclusive tactile picture books by digital fabrication and generative AI,'' in \emph{ACM SIGGRAPH 2023 Labs}, 2023, doi: 10.1145/3588029.3595471.

\bibitem{tanaka2025} K. Tanaka, A. Tsutsui, and Y. Ochiai, ``A comparative evaluation of GenAI-assisted tactile graphics from the perspective of totally blind professionals in tactile graphic supervision,'' in \emph{Proc. ASSETS}, 2025, pp. 1-5, doi: 10.1145/3663547.3759740.

\bibitem{chhedakothary2026} A. Chheda-Kothary, J. E. Froehlich, and J. O. Wobbrock, ``Opportunities for AI-enhanced visual story creation by blind and low-vision children,'' in \emph{Proc. ASSETS}, 2026, doi: 10.1145/3797867.3829063.

\bibitem{jayant2007} C. Jayant, M. Renzelmann, D. Wen, S. Krisnandi, R. Ladner, and D. Comden, ``Automated tactile graphics translation: In the field,'' in \emph{Proc. ASSETS}, 2007, pp. 75-82.

\bibitem{zolek2024} K. Żołek-Tryznowska, A. Brzezińska, and A. Bednarczyk, ``Quality assessment of braille dots printed by fused deposition modeling 3D printing technology,'' \emph{Advances in Science and Technology Research Journal}, 2024, pp. 321-330, doi: 10.12913/22998624/191928.

\bibitem{puerta2024} E. Puerta, T. Crnovrsanin, L. South, and C. Dunne, ``The effect of orientation on the readability and comfort of 3D-printed braille,'' in \emph{Proc. ACM CHI Conf. Human Factors in Computing Systems}, 2024, pp. 1-15, doi: 10.1145/3613904.3642719.

\bibitem{iso17049} International Organization for Standardization, ``Accessible design: Application of braille on signage, equipment and appliances,'' ISO 17049:2013, 2013.

\bibitem{ueb} International Council on English Braille, ``The Rules of Unified English Braille,'' 2024.

\bibitem{liblouis} Liblouis Project, ``Liblouis: An open-source braille translator and back-translator,'' release 3.39.0. [Online]. Available: https://liblouis.io

\bibitem{wcag22} World Wide Web Consortium, ``Web Content Accessibility Guidelines (WCAG) 2.2,'' W3C Recommendation, 2023.

\bibitem{webllm} C. F. Ruan \emph{et al.}, ``WebLLM: A high-performance in-browser LLM inference engine,'' arXiv:2412.15803, 2024.

\bibitem{tattodev} F. Giuliani, ``tatto.dev,'' 2026. [Online]. Available: https://github.com/francescogiuliani87/tatto.dev

\bibitem{tactv2} I. Fasolino, ``Tact: A zero-cost, browser-based pipeline for on-demand tactile braille storybooks,'' 2026, doi: 10.5281/zenodo.22698645, arXiv:2609.12272.

\end{thebibliography}
\end{document}